\pdfoutput=1

\documentclass[11pt,twoside,a4paper,cmspaper,final,collab]{cms-tdr}
\def\svnVersion{490885P}\def\cmsCernNoTag{CERN-EP-2018-247}\def\cmsCernDate{\today}\def\cmsMessage{Published in Physical Review D as \href{http://dx.doi.org/10.1103/PhysRevD.99.012010}{\doi{10.1103/PhysRevD.99.012010.}}}

\begin{document}\cmsNoteHeader{EXO-17-030}

\hyphenation{had-ron-i-za-tion}
\hyphenation{cal-or-i-me-ter}
\hyphenation{de-vices}
\RCS$Revision: 485372 $
\RCS$HeadURL: svn+ssh://svn.cern.ch/reps/tdr2/papers/EXO-17-030/trunk/EXO-17-030.tex $
\RCS$Id: EXO-17-030.tex 485372 2018-12-20 23:29:11Z alverson $

\newlength\cmsFigWidth
\ifthenelse{\boolean{cms@external}}{\setlength\cmsFigWidth{0.98\columnwidth}}{\setlength\cmsFigWidth{0.75\textwidth}}
\ifthenelse{\boolean{cms@external}}{\providecommand{\cmsLeft}{upper\xspace}}{\providecommand{\cmsLeft}{left\xspace}}
\ifthenelse{\boolean{cms@external}}{\providecommand{\cmsRight}{lower\xspace}}{\providecommand{\cmsRight}{right\xspace}}
\newlength\cmsTabSkip\setlength{\cmsTabSkip}{1ex}
\ifthenelse{\boolean{cms@external}}{\providecommand{\cmsTable}[1]{#1}}{\providecommand{\cmsTable}[1]{\resizebox{\textwidth}{!}{#1}}}

\cmsNoteHeader{EXO-17-030}
\title{Search for pair-produced three-jet resonances in proton-proton collisions at \texorpdfstring{$\sqrt{s}=13$\TeV}{sqrt(s) = 13 TeV}}

\date{\today}

\abstract{
         A search has been performed for pair-produced resonances decaying into three jets. The proton-proton collision data used for this analysis were collected with the CMS detector in 2016 at a center-of-mass energy of 13\TeV and correspond to an integrated luminosity of 35.9\fbinv. The mass range from 200 to 2000\GeV is explored in four separate mass regions. The observations show agreement with standard model expectations. The results are interpreted within  the framework of $R$-parity violating SUSY, where pair-produced gluinos decay to a six quark  final state. Gluino masses below 1500\GeV are excluded at 95\% confidence level. An analysis based on data with multijet events reconstructed at the trigger level extends the reach to masses as low as 200\GeV. Improved analysis techniques have led to enhanced sensitivity, allowing the most stringent limits to date to be set on gluino pair production.
}

\hypersetup{%
pdfauthor={CMS Collaboration},%
pdftitle={Search for pair-produced three-jet resonances in proton-proton collisions at sqrt(s) =13 TeV},%
pdfsubject={CMS},%
pdfkeywords={CMS, physics, multijets, SUSY}}

\maketitle
\section{Introduction}
Multijet final states at hadron colliders provide a unique window into many possible extensions of the standard model (SM), albeit in the presence of large SM background processes. Many of these models predict resonances, such as heavy colored fermions transforming as octets under $SU(3)_\textrm{c}$ \cite{tcgut,exoquarks,unichiral,altchiral} or supersymmetric gluinos that undergo $R$-parity violating (RPV) decay into three quarks~\cite{sixjets,multijets,reuven}. All analyses of data collected at the Fermilab Tevatron by CDF~\cite{cdfmultijets} and at Run 1 of the CERN LHC by CMS~\cite{gluino2011,gluino2012} at $\sqrt{s}=7$ and $8$\TeV   used the jet-ensemble method to suppress the large SM background. Searches for similar signals have been performed by ATLAS~\cite{atlas7,atlas8, atlas13} at $\sqrt{s}=7$, 8 and 13\TeV. These analyses provide limits that exclude gluinos undergoing RPV decays, for gluino masses below 144, 650, and 917\GeV for the Tevatron, CMS, and ATLAS results, respectively.

Presented here are the results of a dedicated search for pair-produced resonances, each decaying into three quarks (referred to as ``three-jet resonances" hereafter) in multijet events in proton-proton (\Pp\Pp) collisions. The study is based on a data sample of $\Pp\Pp$ collisions at $\sqrt{s}=13$\TeV, corresponding to an integrated luminosity of $35.9\pm0.9$\fbinv~\cite{lumi}, collected in 2016 with the CMS detector~\cite{Chatrchyan:2008zzk}. Events with at least six jets, each with high transverse momentum (\pt), are selected and investigated for the presence of three-jet resonances consistent with strongly coupled particle decays. The event selection criteria are optimized using a supersymmetric gluino model with the assumption that the gluinos decay with a 100\% branching fraction to quarks. Compared to previous analyses, this search extends its reach to lower masses because of improvements in data acquisition. Additionally, improvements in analysis techniques such as use of Dalitz variables and new selection algorithms significantly enhance sensitivity over the entire mass spectrum. We observe an improvement in sensitivity by a factor of  6.2 (1.8) at 200 (2000)\GeV compared to the previous best limits.

\section{The CMS detector}

The central feature of the CMS apparatus~\cite{Chatrchyan:2008zzk} is a superconducting solenoid of 6\unit{m} internal diameter, providing a magnetic field of 3.8\unit{T}. Within the solenoid volume are a silicon pixel and strip tracker, a lead tungstate crystal electromagnetic calorimeter (ECAL), and a brass and scintillator hadron calorimeter (HCAL), each composed of a barrel and two endcap sections. Forward calorimeters extend the pseudorapidity ($\eta$) coverage provided by the barrel and endcap detectors. Muons are detected in gas-ionization chambers embedded in the steel flux-return yoke outside the solenoid. A particle-flow (PF) algorithm~\cite{CMS-PRF-14-001} aims to reconstruct and identify each individual particle in an event, with an optimized combination of information from the various elements of the CMS detector. The energy of photons is obtained from the ECAL measurement. The energy of electrons is determined from a combination of the electron momentum at the primary interaction vertex as determined by the tracker, the energy of the corresponding ECAL cluster, and the energy sum of all bremsstrahlung photons spatially compatible with originating from the electron track. The energy of muons is obtained from the curvature of the corresponding track. The energy of charged hadrons is determined from a combination of their momentum measured in the tracker and the matching ECAL and HCAL energy deposits, corrected for zero-suppression effects and for the response function of the calorimeters to hadronic showers. Finally, the energy of neutral hadrons is obtained from the corresponding corrected ECAL and HCAL energy. The physics objects are the jets, clustered with the tracks assigned to the vertex as inputs, and the associated missing transverse momentum, taken as the negative vector sum of the \pt of those jets. The reconstructed vertex with the largest value of summed physics-object $\pt^2$ is taken to be the primary $\Pp\Pp$ interaction vertex.

Jets are reconstructed from the energy deposits in the calorimeter towers together with the tracks assigned to the vertex, clustered using the anti-\kt algorithm~\cite{Cacciari:2008gp, Cacciari:2011ma} with a distance parameter of 0.4 (referred as AK4 jets). Jet momentum is determined as the vectorial sum of all particle momenta in the jet, and is found from simulation to be within 5 to 10\% of the true momentum over the whole \pt spectrum and detector acceptance. Additional proton-proton interactions within the same or nearby bunch crossings can contribute additional tracks and calorimetric energy depositions to the jet momentum. To mitigate this effect, tracks identified to be originating from pileup vertices are discarded, and an offset correction is applied to correct for remaining contributions. Jet energy corrections are derived from simulation to bring the measured response of jets to that of particle level jets on average. In situ measurements of the momentum balance in dijet, photon+jet, {\PZ}+jet, and multijet events are used to account for any residual differences in jet energy scale in data and simulation~\cite{Khachatryan:2016kdb}. Additional selection criteria are applied to each jet to remove jets potentially dominated by anomalous contributions from various subdetector components or reconstruction failures~\cite{CMS:2017wyc}. The jet energy resolution amounts typically to 15\% at 10\GeV, 8\% at 100\GeV, and 4\% at 1\TeV.

\section{Triggers}

Events of interest are selected using a two-tiered trigger system~\cite{Khachatryan:2016bia}. The first level (L1), composed of custom hardware processors, uses information from the calorimeters and muon detectors to select events at an average rate of around 100\unit{kHz} within a time interval of less than 4\mus. The second level, known as the high-level trigger (HLT), consists of a farm of processors running a version of the full event reconstruction software optimized for fast processing, and reduces the event rate to around 1\unit{kHz} before data storage. To keep the recorded data rate low, high thresholds are imposed for the triggers used to study jet-based physics, such as requiring high-\pt jets and a large \HT (scalar sum of AK4 jet \pt values).

For the high-mass search, covering the signal mass region above 700\GeV, we use events collected by the OR of two different triggers: the first requires $\HT\geq 800\GeV$ calculated with jet $\pt \geq 40\GeV$; and the second requires at least four jets with $\pt \geq 70\GeV$ and $\HT \geq750\GeV$. Hereafter this set of triggers will be referred to as jets+\HT. In order to achieve full trigger efficiency for events passing the offline selection, the following selection is imposed: $\HT \geq900\GeV$ with jet $\pt \geq50\GeV$ and jet multiplicity ($N_\text{jets}$) $\geq6$. All jets are required to be within $\abs{\eta}<2.4$. The high thresholds of this trigger makes it insensitive to physics at low mass scales (${\sim} 200\GeV$).

To probe new physics at low mass scales, the selection criteria for the trigger must be relaxed. The trigger used for the low-mass search is called the PF Scouting trigger, which has an \HT requirement of $\geq410\GeV$ calculated with jet $\pt \geq 20\GeV$. This results in an event record rate about 2\unit{kHz}. Owing to limitation on the available bandwidth, a minimal amount of information is stored per event, specifically: PF objects, comprising jets, leptons and photons as reconstructed at the HLT. This yields an event size of 10\unit{KB}/event which, is significantly smaller than the 1\unit{MB} event size for normal triggers. The thresholds of the PF Scouting trigger allow us to reconstruct the fully hadronic decay of the top quark, which provides a well understood three-jet resonance signal to validate both the PF Scouting trigger and the
search strategy. In order to achieve full trigger efficiency for events passing the offline selection, the following selection is imposed: $\HT \geq 650\GeV$ with jet $\pt \geq 30\GeV$ and $N_\text{jets}$ $\geq6$. All jets are required to be reconstructed within $\abs{\eta}<2.4$.

\section{Generation of simulated events}

Pair-produced gluinos are used to model the signal. Gluino production is simulated using \MGvATNLO 2.2.2 \cite{madgraph} and gluino decays are simulated using \PYTHIA 8.212 \cite{pythia}, with each gluino decaying into three jets via the $\lambda_{\PQu\PQd\PQd}$ quark RPV coupling. The coupling is set such that the branching fraction of the gluino to three jets is 100\%. The masses of the generated gluinos range from 200 to 2000\GeV in steps of 100\GeV. For the generation of this signal, all superpartners except the gluino are decoupled~\cite{reuven} by setting the squark masses to high values. The natural width of the gluino resonance is assumed to be much smaller than the resolution of the detector, and no intermediate particles are produced in the gluino decay. Simulation of the CMS detector is performed using \GEANTfour~\cite{geant4}.

All simulated samples are produced with the parton distribution functions (PDF) NNPDF3.0~\cite{Ball:2014uwa}, with the precision (LO or NLO) set by the generator used.

\section{Event selection}
\label{sec:evtsel}

Events, recorded with the PF Scouting and jets+\HT triggers described above, are required to have at least one reconstructed primary vertex~\cite{goodPV}. Since this analysis targets pair-produced three-jet resonances, we require events to contain at least six reconstructed jets.

To identify the three jets (triplet) produced by gluino decay in these multijet events, we extend the jet ensemble technique~\cite{cdfmultijets, cmsmultijets} by examining the internal dynamics of multijet events. This technique examines all possible triplets in each multijet event and applies selection criteria to the events, pairs of triplets, and individual triplets to maximize signal sensitivity. We find that restricting the set of considered triplets to the ones involving only the six jets of highest \pt in events with more than six jets, maximizes our sensitivity to the signal, while keeping the background manageable. From the combinatorics of 3 jets chosen from an ensemble of 6, we reconstruct 20 triplets per event, corresponding to two pairs of 10 triplets. For signal events, at most two triplets come from the pair-produced gluino decay, with the remaining triplets corresponding to incorrect jet combinations.

After the offline selection requirements mentioned above, we impose further selection criteria in two steps. In the first, we apply a selection based on event-level variables exploiting the kinematic features and decay topology of the event as a whole. In the second step, we impose selection requirements on variables defined by the features of the triplets and triplet pairs.

\subsection*{Dalitz Variables}

\begin{figure}
\centering
   \includegraphics[width=0.45\textwidth]{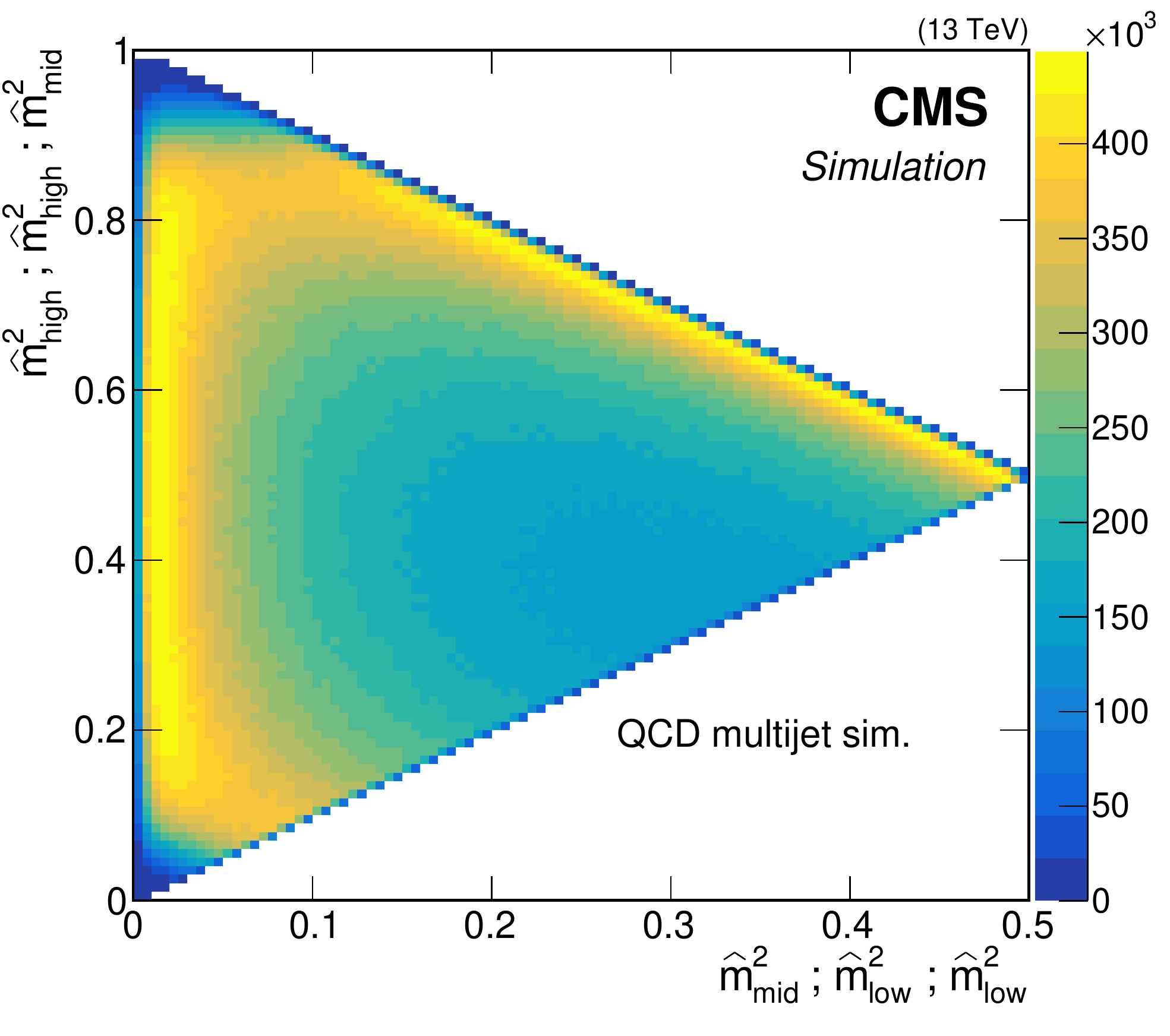}
   \includegraphics[width=0.45\textwidth]{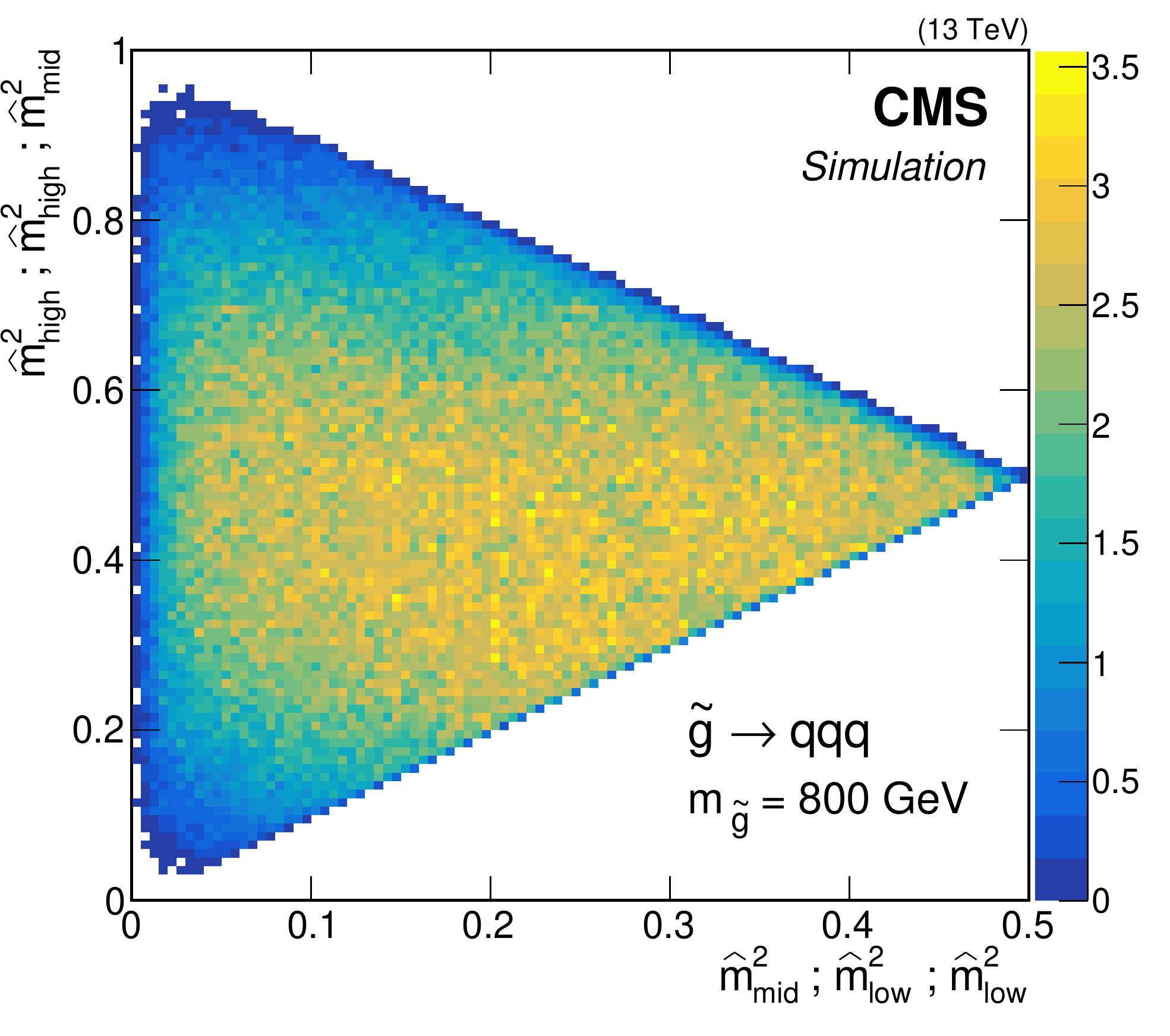}
   \caption{Pair masses within the triplet as described in Eq.~(\ref{mhat32}) plotting superimposed $\hat{m}(3,2)^2_{\text{high}}$ vs. $\hat{m}(3,2)^2_{\text{low}}$, $\hat{m}(3,2)^2_{\text{high}}$ vs. $\hat{m}(3,2)^2_{\text{mid}}$ and $\hat{m}(3,2)^2_{\text{mid}}$ vs. $\hat{m}(3,2)^2_{\text{low}}$. QCD multijet triplets (\cmsLeft) cluster at the edge, while triplets from  signal events  ($m_{\tilde{g}}$ = 800\GeV, \cmsRight) fill the center.}
 \label{fig:dalitz}
\end{figure}

A very useful technique for studying three-body decays uses Dalitz plots, developed by R.H. Dalitz to study K meson decays \cite{dalitz}. Dalitz plots are used to study internal resonances in three body decays. We extend this formalism to construct Dalitz variables that contain information about the internal dynamics of the three-body decay, in order to differentiate between the gluino decays and QCD multijet backgrounds. To construct the Dalitz variables, we form the invariant masses of three dijet pairs inside the triplet, with masses $m_{12}$, $m_{23}$, $m_{13}$. Dalitz variables for a triplet are formed by normalizing these dijet invariant masses. They are defined as follows
\begin{equation}
\label{mhat32}
\hat{m}(3,2)^2_{ij} = \frac{m_{ij}^2}{m_{ijk}^2+m_{i}^2+m_{j}^2+m_{k}^2}.
\end{equation}
Here, $m_{i}$ are the mass of the individual jets and $m_{ijk}$ is the mass of the triplet. Indicies here refer to jets in the triplet, where $i,j,k \in \{1,2,3\}$. There are three $\hat{m}(3,2)^2_{ij}$ in a triplet; we express this with the label ``(3,2)'' where the ``3'' refers to the overall object being a triplet and the ``2'' refers to pairs inside this triplet. The invariant mass of the dijet pairs is normalized such that their Dalitz variables sum up to unity and are dimensionless. For signal triplets, the lack of an internal resonance and the evenly spread out jets make the Dalitz variables close to the value 1/3, implying a symmetric decay, where the jets have uniform geometric separation in the center-of-mass frame of the gluino. Triplets made of jets arising from QCD multijets are more asymmetric, resulting in their $\hat{m}(3,2)^2_{ij}$ being closer to 0 or 1. This is illustrated in Fig. \ref{fig:dalitz}. The three $\hat{m}(3,2)^2_{ij}$ values per triplet are sorted from largest to smallest, and labeled $\hat{m}(3,2)^2_{\text{high}}$, $\hat{m}(3,2)^2_{\text{mid}}$,  and $\hat{m}(3,2)^2_{\text{low}}$. We plot the three pairs of these $\hat{m}(3,2)^2$s per event: $\hat{m}(3,2)^2_{\text{high}}$ vs $\hat{m}(3,2)^2_{\text{mid}}$, $\hat{m}(3,2)^2_{\text{high}}$ vs $\hat{m}(3,2)^2_{\text{low}}$, and $\hat{m}(3,2)^2_{\text{mid}}$ vs $\hat{m}(3,2)^2_{\text{low}}$. These three pairs occupy mutually exclusive regions in the $\hat{m}(3,2)^2$ vs $\hat{m}(3,2)^2$ plane, which combine to give a single overall distribution. This plot is referred to as a dimensionless Dalitz plot. When the variables are displayed in a Dalitz plot, the signal peaks in the center closer to the value 1/3 while the QCD multijet background clusters around the edges.

\begin{figure}
\centering
   \includegraphics[width=0.48\textwidth]{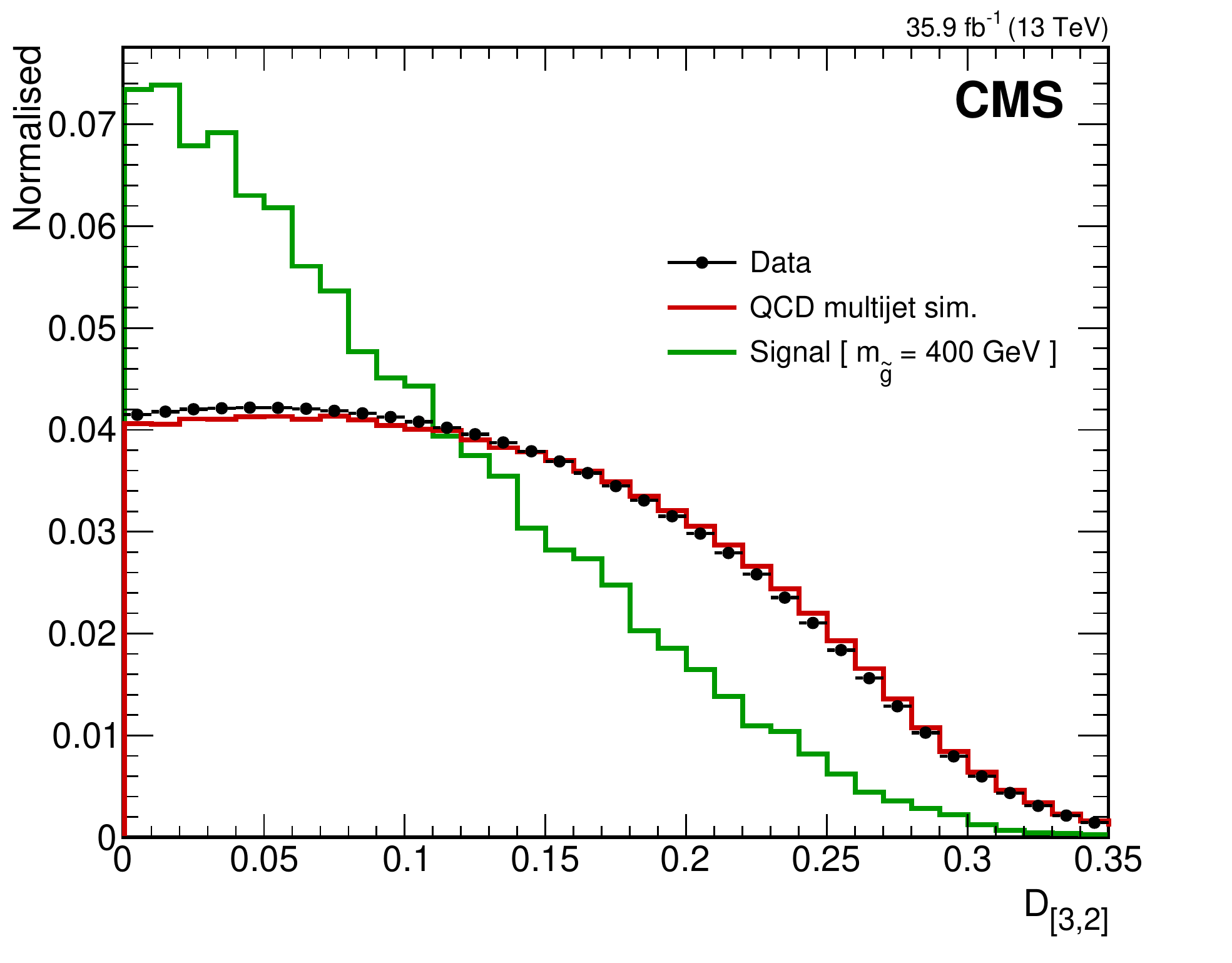}
   \includegraphics[width=0.48\textwidth]{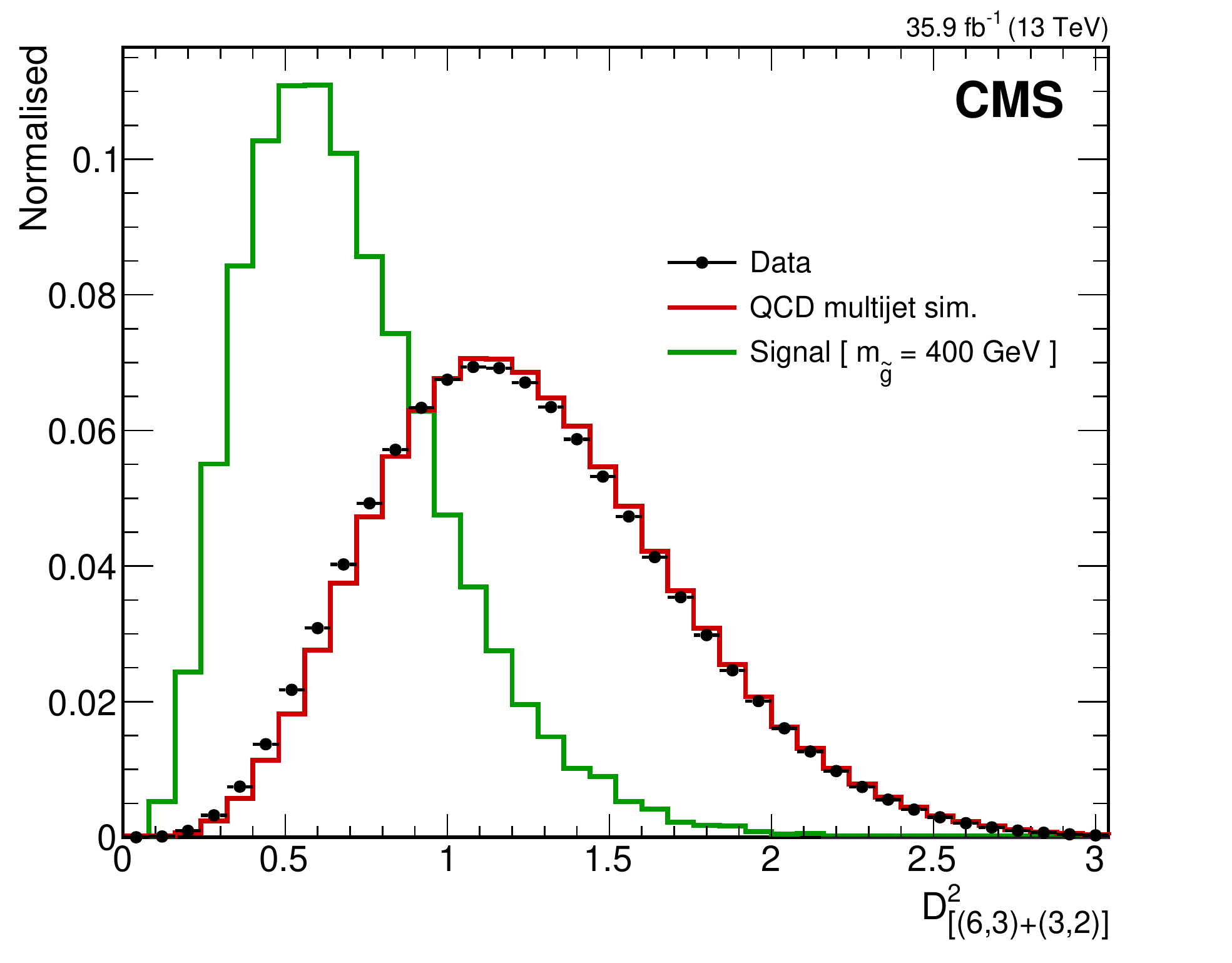}
   \caption{\cmsRight: The $D^2_{[3,2]}$ variable as described in Eq.~(\ref{eq:mds32}) for signal (gluino of mass 400\GeV) and QCD multijet triplets. \cmsLeft: The $D^2_{[(6,3)+(3,2)]}$ distribution as described in Eq.~(\ref{eq:mds6332}), for signal (gluino of mass 400\GeV) and QCD multijet triplets. The distributions are made after nominal selection criteria.}
 \label{fig:mds32}
\end{figure}

Using this feature, we define a variable called mass distance squared (or $D^2$) to characterize the symmetry between the jets inside a triplet. This variable, which is plotted in Fig. \ref{fig:mds32}, is defined as
\begin{equation}
\label{eq:mds32}
D^2_{[3,2]}=\sum_{i>j} \left(\hat{m}(3,2)_{ij}-\frac{1}{\sqrt{3}}\right)^2 .
\end{equation}
We extend this idea to the event-level to define a second variable, to estimate the angular spread of the 6 jets within a pairs of triplets. This distance measure will have a low value for signal-like topologies, indicating well separated jets with similar momentum and a high value for dijet-like topologies such as QCD. For this purpose, new Dalitz variables are defined as normalized invariant mass of jet triplets constructed from the six highest \pt jets
\begin{equation}
\hat{m}(6,3)^2_{i j k}= \frac{m^2_{i j k}}{4~ m^2_{i j k l m n}+6~ \sum_i m^2_i}.
\end{equation}
Here, $m_{i j k l m n}$ is the invariant mass of the top six jets, ordered in \pt. Indicies here refer to the the top six jets ordered in \pt, where $i,j,k,l,m,n \in \{1,2,...,6 \}$. The label ``(6,3)'', the first index refers to the overall object being a six-jet ensemble, and the second refers to triplets inside this six-jet ensemble. For a six-jet topology, we will have 20 such $\hat{m}(6,3)^2_{i j k}$ variables.  Six-jet events from QCD multijets are largely due to a core dijet event with extra radiated jets.  These jets tend to be grouped together.  Jets from pair-produced gluino decays tend to be distributed more uniformly across the detector. This makes these variables close to 0 or 1 for QCD multijets and close to $1/20$ for jets coming from signal decay. The invariant mass of the triplet is normalized such that these 20 event-level Dalitz variables sum up to 1. Using the previously defined variables, we define the following six-jet distance measure in a similar way to $D^2_{[3,2]}$,
\begin{equation}
\label{eq:mds6332}
D^2_{[(6,3)+(3,2)]} = \sum_{i<j<k} \left( \sqrt{\hat{m}(6,3)^2_{i j k}+D^2_{[3,2],ijk}} - \frac{1}{\sqrt{20}} \right)^2.
\end{equation}
This $D^2_{[(6,3)+(3,2)]}$ combines the $D^2_{[6,3]}$ and $D^2_{[3,2]}$ into a single event-level variable. Figure \ref{fig:mds32} shows the $D^2_{[(6,3)+(3,2)]}$ and $D^2_{[3,2]}$ distributions for QCD multijet background and gluino simulation after the selection criteria: $\HT \geq 650\GeV$, sixth jet $\pt \geq 50\GeV$ and $N_\text{jets}$ $\geq 6$. The small disagreement between QCD multijet simulation and data visible in Fig. \ref{fig:mds32} is due to imperfect modeling of the QCD multijet simulation. Since the QCD multijet simulation is not used for predicting the background, this discrepancy has a negligible effect on this search.”

\subsection*{Other pair and triplet level selections}

For each triplet pair, we calculate a variable called ``mass asymmetry", defined as
\begin{equation}
A_m =\frac{\abs{m_{ijk}-m_{lmn}}}{m_{ijk}+m_{lmn}}.
\end{equation}
Here, $m_{ijk}$ and $m_{lmn}$ are masses of the two unique triplets in a triplet pair. This variable shows discriminating power between signal and background.

\begin{figure}
\centering
   \includegraphics[width=0.48\textwidth]{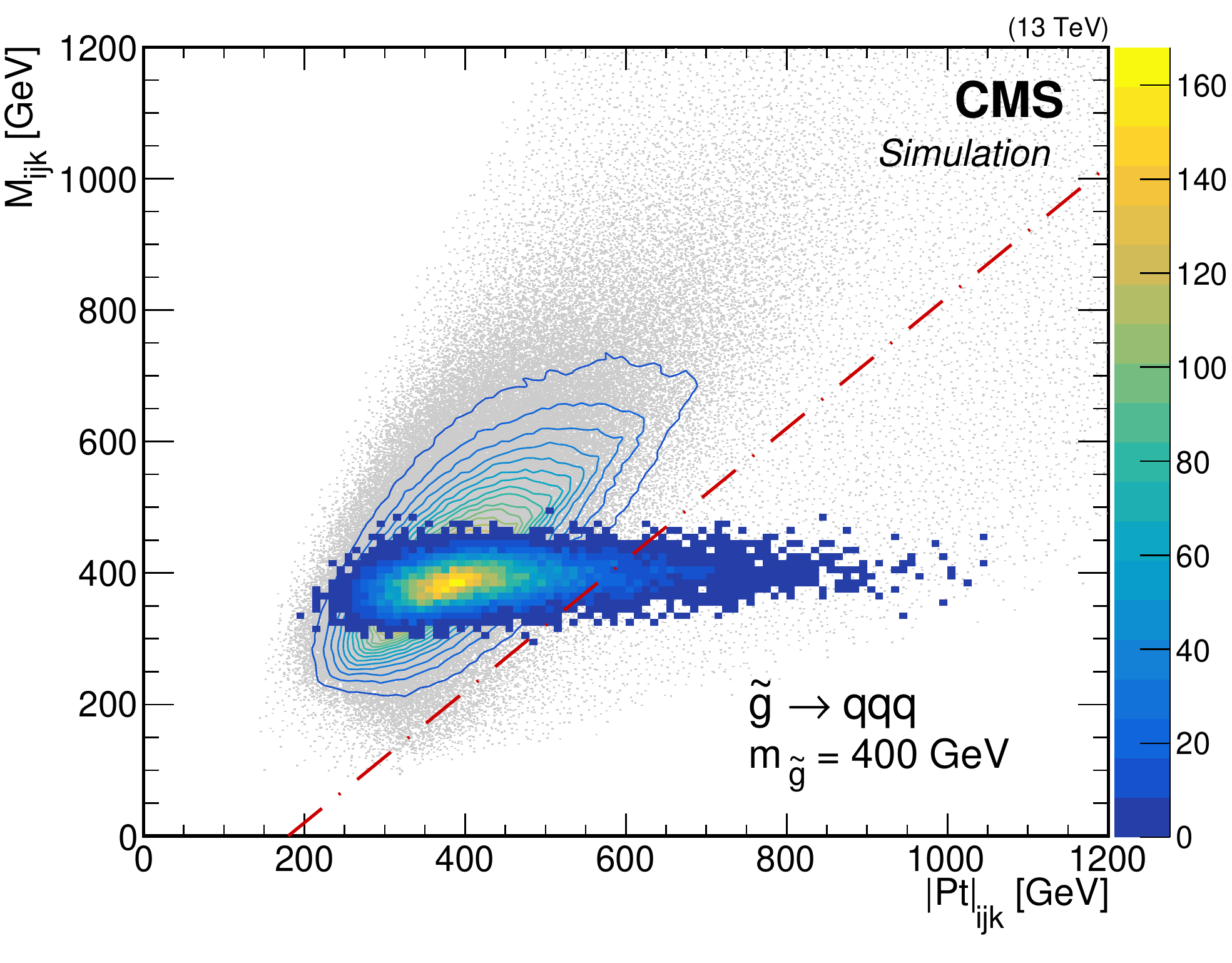}
   \caption{Expected distribution of triplet invariant mass versus the scalar sum of jet \pt in the triplet for a gluino of mass 400\GeV decaying into jets. The filled color represents correctly reconstructed signal triplets, while the gray points and contour lines represent the QCD multijet background and combinatorial background from signal events. The red dashed line illustrates the $\Delta$ cut: triplets to the right of the line are accepted by the selection criterion.}
 \label{fig:MjjjvsSumpt}
\end{figure}

For triplets from multijet QCD events or combinatorial background, the scalar sum \pt($\abs{\pt}_{ijk}$) will scale with the triplet mass ($m_{ijk}$). Whereas it is not case for signal triplets as they have constant invariant mass. We exploit this feature of signal triplets by constructing a selection, referred to as a``Delta cut", defined as:
\begin{eqnarray}
m_{ijk} \,<\,\abs{\pt}_{ijk}\,-\,\Delta,
\label{eq:diag_def}
\end{eqnarray}
where $m_{ijk}$~is the triplet invariant mass, $\abs{\pt}_{ijk}$ is the scalar sum of jet \pt in the triplet (triplet scalar \pt), and $\Delta$ is an adjustable offset. A scatter plot of the triplet invariant mass versus triplet scalar \pt for a gluino with a mass of 400\GeV is shown in Fig.~\ref{fig:MjjjvsSumpt}, which clearly shows that by imposing this selection criterion we eliminate most of the background while retaining a significant fraction of the signal.

We optimize selection criteria in four separate mass ranges with a metric defined as the ratio of signal to the square root of the background obtained by integrating the triplet mass distribution from gluino and QCD multijet simulations in a window around the signal peak. We note that background can include triplets from QCD multijet, as well as combinatorial background from signal. The four resulting signal regions are defined in Table \ref{table:optpt} and labeled from 1 to 4.

\begin{table*}[htbp]
\topcaption{ Gluino mass ranges used in this analysis, and selection criteria used. Note that the  gluino mass ranges in the upper two rows in the table use events collected using the PF Scouting trigger, while the lower two rows in the table use events collected using jets+\HT trigger. The symbol `$>$' and `$<$' represent the direction of the cut.}
\cmsTable{\begin{scotch}{ccccccccc}
    Region & \begin{tabular}[c]{@{}c@{}}Gluino\\ mass range\end{tabular}  & Jet \pt & \HT & sixth jet \pt & $D^2_{[(6,3)+(3,2)]}$ & $A_m$ & $\Delta$ & $D^2_{[3,2]}$ \\ \hline
    1 & 200--400\GeV & $>$30\GeV& $>$650\GeV& $>$40\GeV& $<$1.25 & $<$0.25 & $>$250\GeV& $<$0.05\\
    2 & 400--700\GeV & $>$30\GeV& $>$650\GeV& $>$50\GeV& $<$1.00 & $<$0.175 & $>$180\GeV& $<$0.175\\
    3 & 700--1200\GeV & $>$50\GeV& $>$900\GeV& $>$125\GeV& $<$0.9 & $<$0.15 & $>$20\GeV& $<$0.2 \\
    4 & 1200--2000\GeV & $>$50\GeV& $>$900\GeV& $>$175\GeV& $<$0.75 & $<$0.15 & $>$-120\GeV& $<$0.25\\
  \end{scotch}}
  \label{table:optpt}
\end{table*}

\section{Background estimation}
\label{sec:sigbkg}

\begin{figure*}
   \centering
   \includegraphics[width=0.48\textwidth]{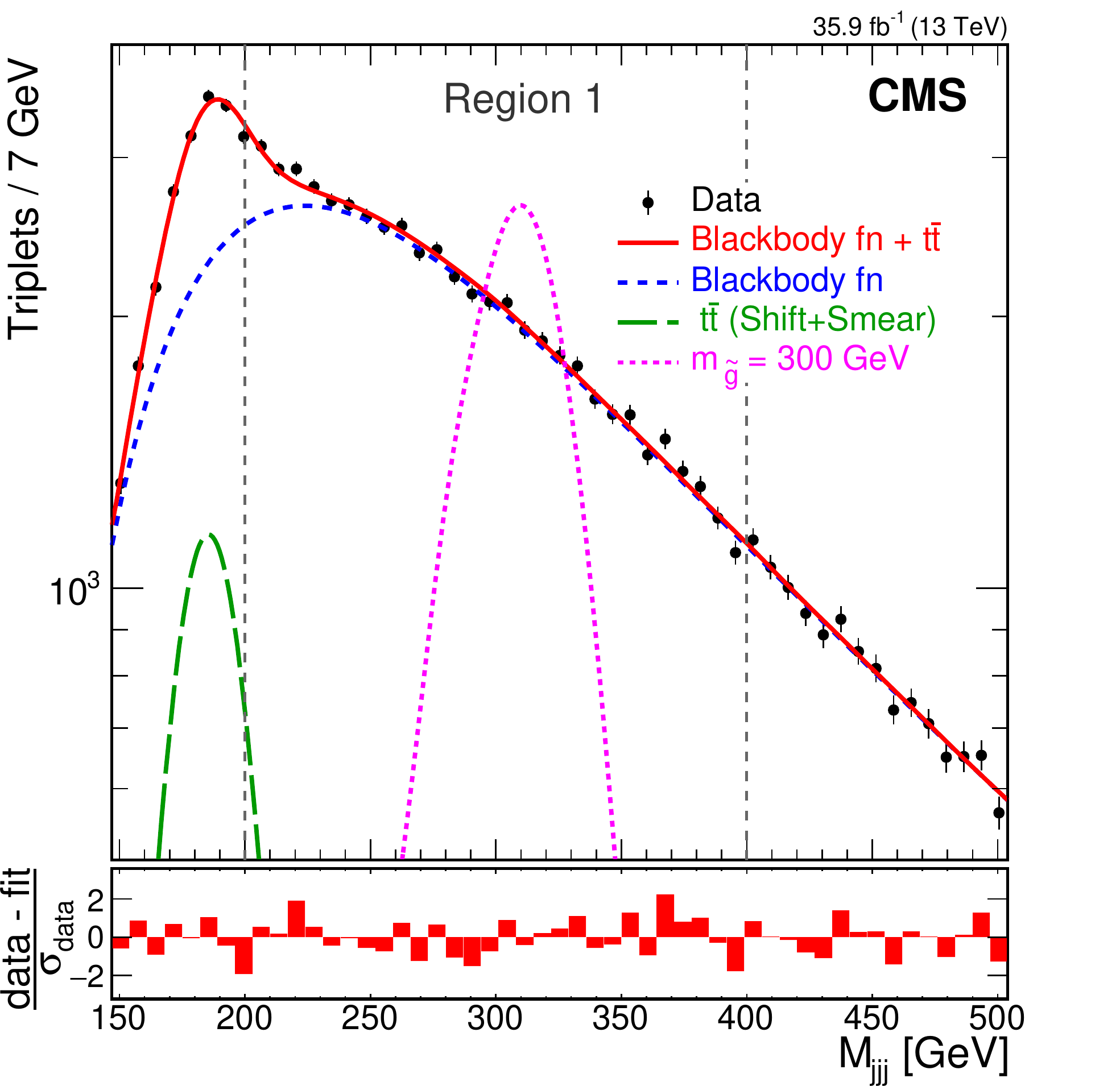}
   \includegraphics[width=0.48\textwidth]{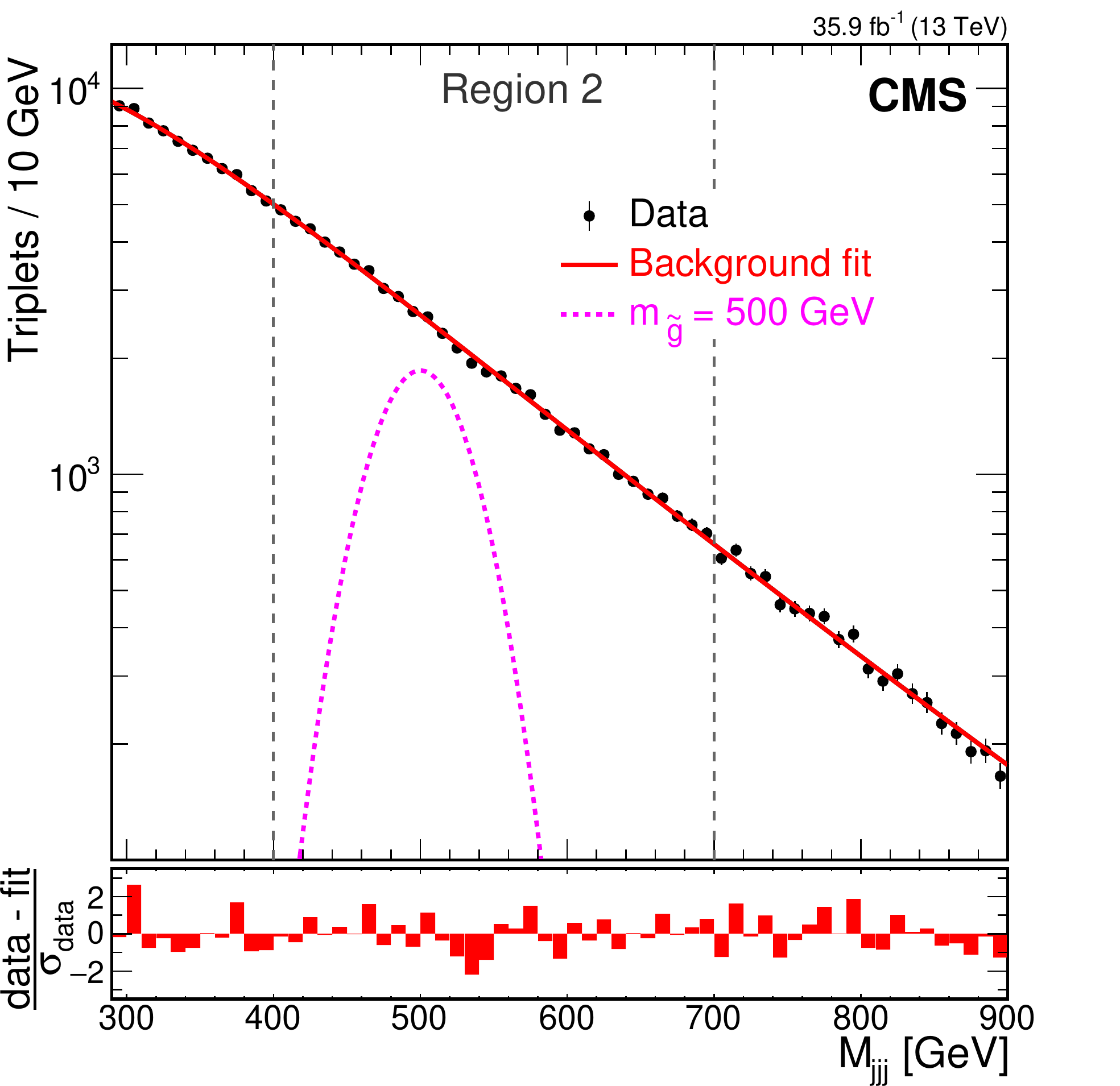}
   \includegraphics[width=0.48\textwidth]{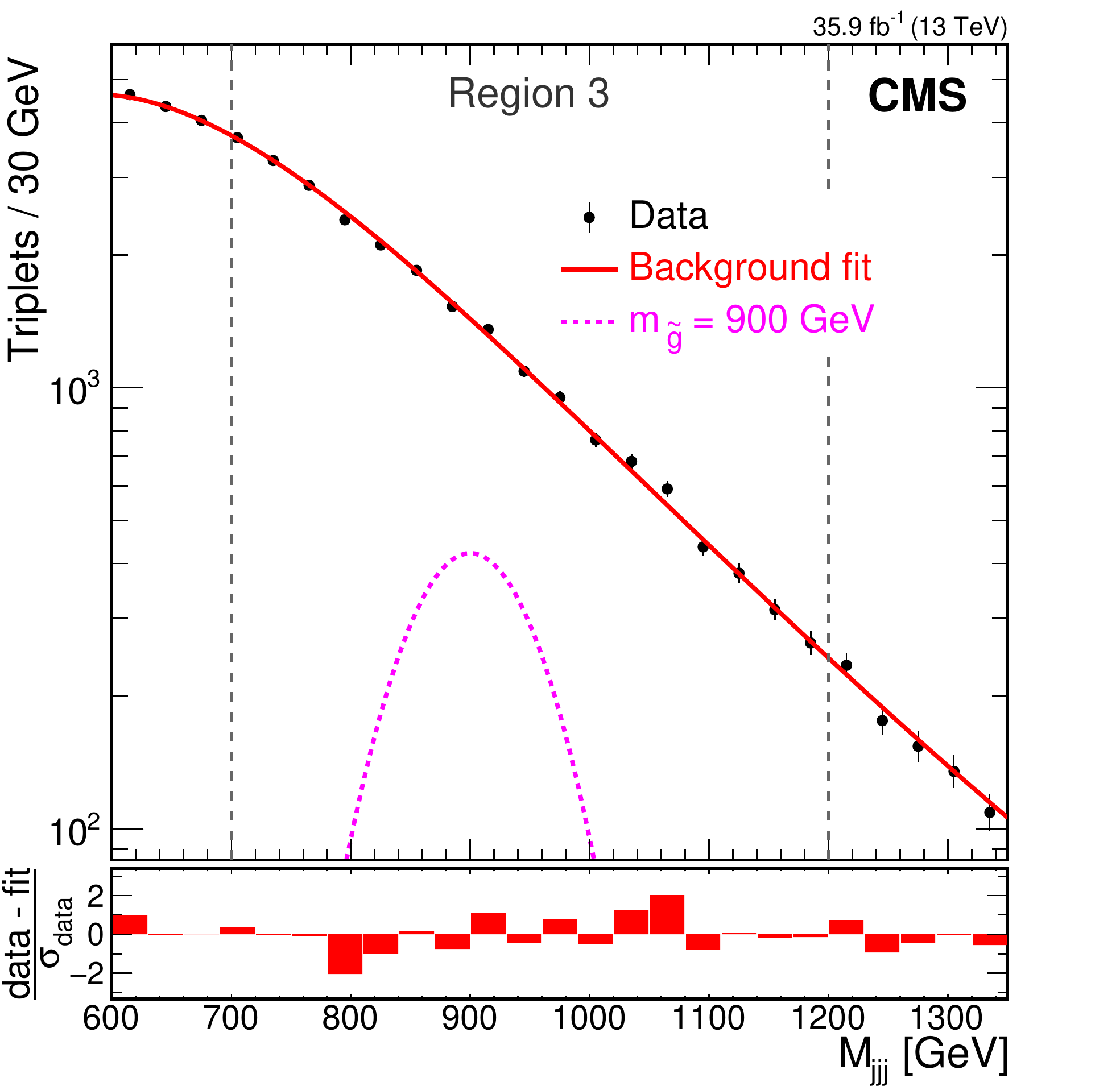}
   \includegraphics[width=0.48\textwidth]{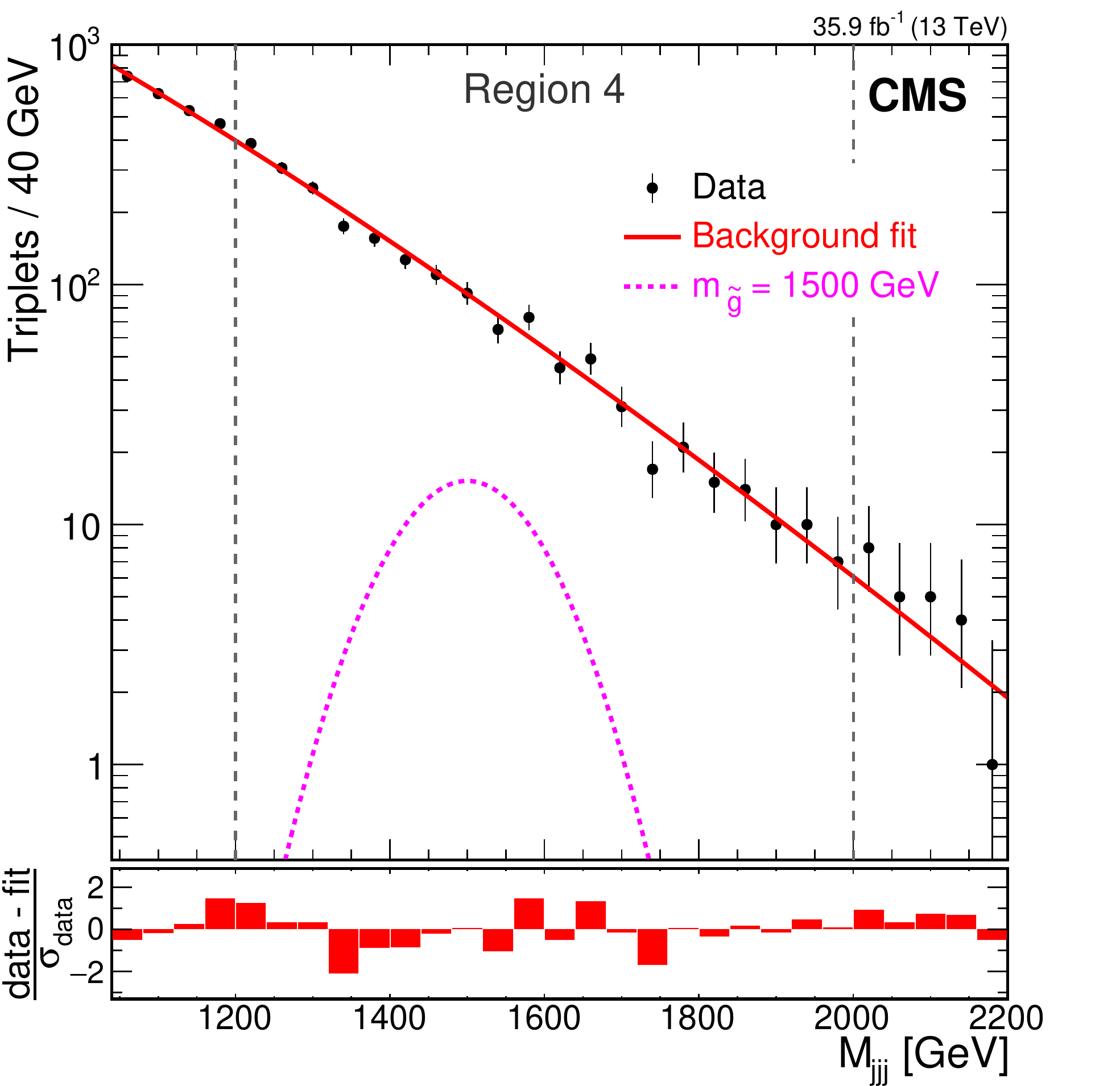}
   \caption{ Mass distributions and background-only fits for each of the mass regions. Region 1 (top left) is fit to a function that combines the blackbody-like term described in Eq.~(\ref{bbf}) with a simulated \ttbar distribution, while the Region 2 and 3 (top right and bottom left) are fit to the four parameter function from Eq.~(\ref{p4}), while the Region 4 (bottom right) is fit to three parameter function from Eq.~(\ref{p4}) with $p_3$ set to zero. The vertical gray lines indicate the mass regions. The gluino signal normalized to the cross section expected from \cite{theoryxscn} is shown in magenta.}
 \label{fig:lf_fits1}
\end{figure*}

There are three sources of background that we consider: QCD multijets, fully hadronic decays of \ttbar pairs, and combinatorial background from signal events. We find that background due to the \ttbar decays is only significant in the lowest mass region of the search. This background is estimated from events simulated with \POWHEG~\cite{powheg1,powheg2,powheg3,powheg4} and their decay is simulated with \PYTHIA. The \ttbar production rate extracted from a background-only fit in Region 1 agrees with the SM expectation within the statistical uncertainty of the measurement. The mass distributions of the QCD multijet and combinatorial backgrounds are estimated by fitting a smooth function to data. Studies of simulated QCD multijet events and combinatorial background from signal events show that the combined mass distribution can be described by a single smooth function. Except for the lowest mass region, the triplet invariant mass background distribution is smoothly falling (as we can see in Fig.~\ref{fig:lf_fits1} ), and we use three types of functions, fit directly to the data, to model this background in different regions.

The background distribution of triplets in Region 1 shows features due to the turn-on of QCD multijet background and \ttbar decays. For modeling this QCD multijet and combinatorial background, we use a function inspired by the formulation of Planck's law of blackbody radiation with an added logarithmic correction to the tails, and this distribution models the background well. This function models the QCD multijet background turn-on better than the four-parameter function, used to fit triplet mass distributions in other regions:
\begin{equation}
\label{bbf}
\frac{\rd N}{\rd x}=\frac{1}{(x+c)^{5+ d \ln{\frac{x}{\sqrt{s}} } }} \frac{a}{\re^{\frac{b}{x+c}}-1},
\end{equation}
where, $a$ is the factor controlling the normalization of the fit, $b$ is the ``temperature" term in blackbody distribution, $c$ controls the translation of the whole distribution, $d$ controls the strength of the logarithmic term, and $\sqrt{s}$ is the center-of-mass energy of the proton-proton collisions.

For modeling the background in Regions 2 and 3, we use the following four-parameter function
\begin{equation}
\label{p4}
\frac{\rd N}{\rd x}=p_0\frac{\left(1-\frac{x}{\sqrt{s}}\right)^{p_1}}{\left(\frac{x}{\sqrt{s}}\right)^{p_2+p_3 \ln{\frac{x}{\sqrt{s}}}}},
\end{equation}
and for Region 4 we used the same parametrization, with $p_{3}$ set to zero, to model the background.

The functional form in Eq.~(\ref{p4}) successfully models the steeply falling dijet mass distribution of QCD multijet production and has been used extensively in dijet resonance searches ~\cite{Khachatryan:2014lpa,Khachatryan:2015dcf}.

We test for possible bias introduced by the choice of background parameterization. We perform signal injection tests on pseudo-experiments generated from QCD multijet simulation. These pseudo-experiments are fit to alternative background parameterizations and the effect on the strength of the extracted signal is examined. These tests indicate a negligible bias. To further validate our background parametrization, we performed pseudo-experiments generated using data from the PF Scouting and jets+\HT event samples. We also perform statistical studies (F-tests) to determine the optimum number of parameters for the background function, to avoid over-constraint. The distributions of triplet mass in the four search regions are shown in Fig.~\ref{fig:lf_fits1} , along with the results of fits to the background-only hypothesis. The mass distributions expected for a typical gluino decay is shown in magenta, with the rate normalized to that expected from \cite{theoryxscn}. The fits reproduce the data distributions well, indicating absence of a signal.

For the signal triplet-mass distribution, signal simulations parameterized with double Gaussian distributions are used. These parameterizations accurately describe the shape of signal triplet mass distribution. The acceptances for the search is defined as the number of correct triplets passing the selection, divided by the number of events generated. The selection criteria given in Table~\ref{table:optpt} result in signal acceptance of $2.6\times10^{-4}$, $8.4 \times 10^{-2}$ and $1.7 \times 10^{-1}$ for the resonance masses $m_{\tilde{g}}= 200$, 900, and 1600\GeV respectively.

\section{Systematic uncertainties}

The search in Regions 1 and 2 uses PF Scouting data. Jets in these events did not have the full offline set of corrections applied. We use the well-measured all-hadronic decay of the top quark to determine the corrections and corresponding systematic uncertainties for the PF Scouting data. The triplet-mass distribution from \ttbar simulation must be adjusted in order to agree with the data in Region 1, with two transformations to the simulated triplet mass distribution required. The first is a translation of 6.6\GeV, referred to as the ``shift'' correction. The second is a convolution with a Gaussian distribution of width of 8.9\GeV, referred to as the ``smear'' correction. The shift and smear values determined from the top resonance measurement are also applied to the gluino simulation. We performed a separate study to investigate the dependence of the shift and smear on the triplet scalar \pt and found negligible correlation. The uncertainties associated with the shift and smear corrections are determined by observing the change in goodness-of-fit metric between simulation and data as these parameters are varied. Corresponding systematic uncertainties for the shift and smear corrections are estimated to be 3.5 and 4\%, respectively. These corrections are defined as a percentage of the mean of the signal distribution. For the jets+\HT data, adjustments analogous to shift and smear are applied to correct for the effects arising from uncertainties in the measurement of jet energy corrections (2.5\%) and jet energy resolution (12\%). These systematic uncertainties affect the shape of the signal triplet mass distributions.

The other systematic uncertainties affecting the yield from the signal samples are the integrated luminosity measurement (2.5\%) and the uncertainty in the determination of acceptance (5\%), which
includes contributions from uncertainties in the PDF. We list the systematic uncertainties for both datasets in Table~\ref{sys_table}.

\begin{table*}[htbp]
\centering
\topcaption{Summary of the systematic uncertainties in the signal yield. For the uncertainty affecting the distribution (shape), the value represents the percentage difference in the nominal value of the systematic uncertainty. These systematic uncertainties are applied to the signal.}
\label{sys_table}
\begin{scotch}{lllc}
Data set                     & Source of systematic  & Effect & Value \\ \hline
\multirow{2}{*}{All}         & Luminosity            & Yield  & 2.5\% \\
                             & Acceptance            & Yield  & 5\%   \\ [\cmsTabSkip]
\multirow{2}{*}{PF Scouting} & Shift                 & Shape  & 3.5\% \\
                             & Smear                 & Shape  & 4\%   \\ [\cmsTabSkip]
\multirow{2}{*}{jets+\HT}    & Jet energy correction & Shape  & 2.5\% \\
                             & Jet energy resolution & Shape  & 12\%  \\
\end{scotch}
\end{table*}

\section{Limits}
The mass distribution of data is described well by the background parameterization, as illustrated in Fig.~\ref{fig:lf_fits1}. We see no significant excess that could indicate the presence of signal, and place upper limits on the product of the cross section and branching fraction for the pair production of three-jet resonances. A modified frequentist approach, with the \CLs criterion as the figure of merit and a profile likelihood as the test statistic, is employed. Limits are calculated with the frequentist asymptotic approximation in RooStats~\cite{Junk:1999kv, Read:2002hq, cowan, RooStats}. The full \CLs calculator gives similar results. The data are fit using a binned maximum-likelihood function, based on the respective four-parameter function. In Region 1, the rate for \ttbar events is set to the value observed from the background-only fit and is allowed to float within the systematic uncertainty. The overall QCD scale is unconstrained and the nuisance parameters effecting the overall rate are introduced as log-normal constrains.

\begin{figure}
\centering
\includegraphics[width=\cmsFigWidth]{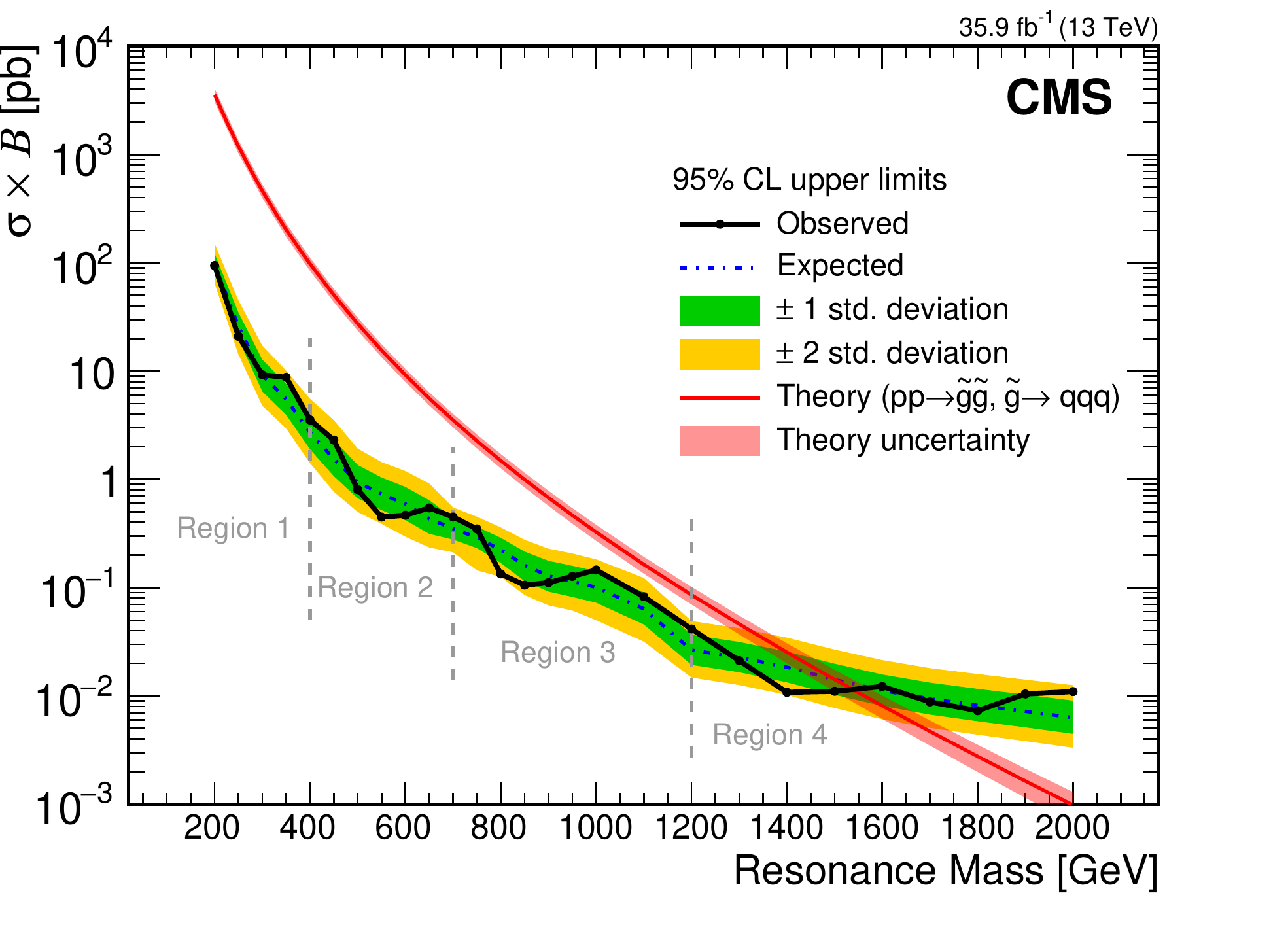}
\caption{Observed and expected frequentist \CLs limits on cross section times branching fraction are calculated in the asymptotic approximation. The solid red curve shows the prediction for the gluino pair productions from~\cite{theoryxscn}. The band around the theory curve indicates the uncertainty associated with PDF and scale choices. The gray vertical lines indicate the boundaries between the mass regions.}
\label{fig:Limits}
\end{figure}

The observed and expected 95\% confidence level (\CL) upper limits on the product of gluino pair-production cross section and branching fraction, as a function of gluino mass, are presented in Fig.~\ref{fig:Limits}. The solid red line in the figure show the next-to-leading order (NLO) plus next-to-leading-logarithm (NLL) cross sections for gluino pair production ~\cite{theoryxscn}, and the shaded region around the solid red line represent the corresponding 1 standard deviation uncertainties, which range from 14 to 31\%. We use the points where the 1 sigma uncertainty curve for the NLO+NLL cross section crosses the observed limit curve to obtain our final results.

The production of gluinos decaying by an $R$-parity violating interaction into jets is excluded at 95\% \CL for gluino masses below 1500\GeV. This is the most stringent mass limit to date on this model of RPV gluino decay, assuming a 100\% branching fraction for gluinos decaying to quark jets.

\section{Summary}

A search has been performed for pair-produced resonances decaying into three jets. The proton-proton collision data used for this analysis were collected with the CMS detector in 2016 at a center-of-mass energy of $\sqrt{s}=13$\TeV and correspond to an integrated luminosity of 35.9\fbinv. The mass range from 200 to 2000\GeV is explored in four separate mass regions. The observations show agreement with standard model expectations. The results are interpreted within the framework of $R$-parity violating SUSY, where pair-produced gluinos decay to a six quark final state. Gluino masses below 1500\GeV are excluded at 95\% confidence level. An analysis based on data with multijet events reconstructed at the trigger level extends the reach to masses as low as 200\GeV. Improved analysis techniques have led to enhanced sensitivity, allowing the most stringent limits to date to be set on gluino pair production.

\begin{acknowledgments}
We congratulate our colleagues in the CERN accelerator departments for the excellent performance of the LHC and thank the technical and administrative staffs at CERN and at other CMS institutes for their contributions to the success of the CMS effort. In addition, we gratefully acknowledge the computing centers and personnel of the Worldwide LHC Computing Grid for delivering so effectively the computing infrastructure essential to our analyses. Finally, we acknowledge the enduring support for the construction and operation of the LHC and the CMS detector provided by the following funding agencies: BMBWF and FWF (Austria); FNRS and FWO (Belgium); CNPq, CAPES, FAPERJ, FAPERGS, and FAPESP (Brazil); MES (Bulgaria); CERN; CAS, MoST, and NSFC (China); COLCIENCIAS (Colombia); MSES and CSF (Croatia); RPF (Cyprus); SENESCYT (Ecuador); MoER, ERC IUT, and ERDF (Estonia); Academy of Finland, MEC, and HIP (Finland); CEA and CNRS/IN2P3 (France); BMBF, DFG, and HGF (Germany); GSRT (Greece); NKFIA (Hungary); DAE and DST (India); IPM (Iran); SFI (Ireland); INFN (Italy); MSIP and NRF (Republic of Korea); MES (Latvia); LAS (Lithuania); MOE and UM (Malaysia); BUAP, CINVESTAV, CONACYT, LNS, SEP, and UASLP-FAI (Mexico); MOS (Montenegro); MBIE (New Zealand); PAEC (Pakistan); MSHE and NSC (Poland); FCT (Portugal); JINR (Dubna); MON, RosAtom, RAS, RFBR, and NRC KI (Russia); MESTD (Serbia); SEIDI, CPAN, PCTI, and FEDER (Spain); MOSTR (Sri Lanka); Swiss Funding Agencies (Switzerland); MST (Taipei); ThEPCenter, IPST, STAR, and NSTDA (Thailand); TUBITAK and TAEK (Turkey); NASU and SFFR (Ukraine); STFC (United Kingdom); DOE and NSF (USA).

\hyphenation{Rachada-pisek} Individuals have received support from the Marie-Curie program and the European Research Council and Horizon 2020 Grant, contract No. 675440 (European Union); the Leventis Foundation; the A. P. Sloan Foundation; the Alexander von Humboldt Foundation; the Belgian Federal Science Policy Office; the Fonds pour la Formation \`a la Recherche dans l'Industrie et dans l'Agriculture (FRIA-Belgium); the Agentschap voor Innovatie door Wetenschap en Technologie (IWT-Belgium); the F.R.S.-FNRS and FWO (Belgium) under the ``Excellence of Science - EOS" - be.h project n. 30820817; the Ministry of Education, Youth and Sports (MEYS) of the Czech Republic; the Lend\"ulet (``Momentum") Program and the J\'anos Bolyai Research Scholarship of the Hungarian Academy of Sciences, the New National Excellence Program \'UNKP, the NKFIA research grants 123842, 123959, 124845, 124850 and 125105 (Hungary); the Council of Science and Industrial Research, India; the HOMING PLUS program of the Foundation for Polish Science, cofinanced from European Union, Regional Development Fund, the Mobility Plus program of the Ministry of Science and Higher Education, the National Science Center (Poland), contracts Harmonia 2014/14/M/ST2/00428, Opus 2014/13/B/ST2/02543, 2014/15/B/ST2/03998, and 2015/19/B/ST2/02861, Sonata-bis 2012/07/E/ST2/01406; the National Priorities Research Program by Qatar National Research Fund; the Programa Estatal de Fomento de la Investigaci{\'o}n Cient{\'i}fica y T{\'e}cnica de Excelencia Mar\'{\i}a de Maeztu, grant MDM-2015-0509 and the Programa Severo Ochoa del Principado de Asturias; the Thalis and Aristeia programs cofinanced by EU-ESF and the Greek NSRF; the Rachadapisek Sompot Fund for Postdoctoral Fellowship, Chulalongkorn University and the Chulalongkorn Academic into Its 2nd Century Project Advancement Project (Thailand); the Welch Foundation, contract C-1845; and the Weston Havens Foundation (USA).
\end{acknowledgments}

\bibliography{auto_generated}

\cleardoublepage \appendix\section{The CMS Collaboration \label{app:collab}}\begin{sloppypar}\hyphenpenalty=5000\widowpenalty=500\clubpenalty=5000\vskip\cmsinstskip
\textbf{Yerevan Physics Institute, Yerevan, Armenia}\\*[0pt]
A.M.~Sirunyan, A.~Tumasyan
\vskip\cmsinstskip
\textbf{Institut f\"{u}r Hochenergiephysik, Wien, Austria}\\*[0pt]
W.~Adam, F.~Ambrogi, E.~Asilar, T.~Bergauer, J.~Brandstetter, M.~Dragicevic, J.~Er\"{o}, A.~Escalante~Del~Valle, M.~Flechl, R.~Fr\"{u}hwirth\cmsAuthorMark{1}, V.M.~Ghete, J.~Hrubec, M.~Jeitler\cmsAuthorMark{1}, N.~Krammer, I.~Kr\"{a}tschmer, D.~Liko, T.~Madlener, I.~Mikulec, N.~Rad, H.~Rohringer, J.~Schieck\cmsAuthorMark{1}, R.~Sch\"{o}fbeck, M.~Spanring, D.~Spitzbart, A.~Taurok, W.~Waltenberger, J.~Wittmann, C.-E.~Wulz\cmsAuthorMark{1}, M.~Zarucki
% ...
\vskip\cmsinstskip
% ...
1:  Also at Vienna University of Technology, Vienna, Austria\\
% ...
\end{sloppypar}
\end{document}